\documentstyle[twocolumn,psfig]{mn}
\title[The Spaghetti Method]
       {Identification of Moving Groups and Member
	Selection using Hipparcos Data\thanks{Based on
	data from the Hipparcos astrometry satellite.}}

\author[Ronnie Hoogerwerf \& Luis A.\ Aguilar]  
       {Ronnie Hoogerwerf$^1$ and Luis A.\ Aguilar$^2$\\ 
	$^1$Sterrewacht Leiden, Postbus 9513, 
	    2300 RA Leiden, the Netherlands\\
	$^2$Instituto de Astronom\'{\i}a, U.N.A.M., Apartado Postal 877, 
	    22800 Ensenada, Baja California, M\'exico}

\begin{document}
\maketitle
\begin{abstract}
A new method to identify coherent structures in velocity
space --- moving groups --- in astrometric catalogues is presented: the
Spaghetti method. It relies on positions, parallaxes, and proper
motions and is ideally suited to search for moving groups in the
Hipparcos Catalogue. No radial velocity information is required.

The method has been tested extensively on synthetic data, and applied
to the Hipparcos measurements for the Hyades and IC2602 open
clusters. The resulting lists of members agree very well with those
of Perryman et al.\ for the Hyades and of Whiteoak
and Braes for IC2602.
\end{abstract}
\begin{keywords}
Astrometry -- Stars: kinematics -- Open clusters and associations:
general -- Open clusters and associations: individual: the Hyades,
IC2602
\end{keywords}

\section{Introduction}\label{sec:intro}
The kinematical distribution of stars in the Solar neighborhood is far
from smooth. In 1869 Proctor already discovered the existence of
groups of stars in the same region of the sky that share kinematic
properties. These groups are the result of the common motion of young
stars produced by the localized nature of star formation in the
Galaxy. Although most of these groups are confined in
configuration space we define a {\it moving group} as a set of stars
which occupy a small volume in velocity space. The distribution of
these stars in configuration space does not enter in this definition.
Among these groups are gravitationally bound open clusters (e.g., the
Hyades [Perryman et al.\ 1998]), unbound OB associations (e.g., the
Sco~OB2 association [de Geus, de Zeeuw \& Lub 1989]), and the
so-called superclusters (e.g., Eggen 1991, Chereul, Cr\'ez\'e \&
Bienaym\'e 1998, 1999). Observational properties of such groups such as,
the initial mass function (e.g., Claudius \& Grosb{\o}l 1980; Brown,
de Geus \& de Zeeuw 1994; Massey, Johnson \& DeGioia-Eastwood 1995),
the local star formation rate and efficiency (e.g., Williams \& McKee
1997), and the fraction and characteristics of binary and multiple
stars (e.g., Blaauw 1991; Brandner et al.\ 1996; Verschueren, David \&
Brown 1996), serve as important tests for current theories on star
formation. These properties and the calibration of the distance scale
and absolute magnitudes depend sensitively on membership. Therefore,
it is crucial to have detailed knowledge of membership of moving
groups, in particular open clusters and OB associations. As the
velocity dispersions in moving groups are small, typically a few ${\rm
km}~{\rm s}^{-1}$ or less (see e.g., Jones \& Herbig 1979;
Hartmann et al.\ 1986; Mathieu 1986; Tian et al.\ 1996), proper
motions and radial velocities can be used to detect the common space
motion and thus determine membership.

For the majority of the moving group candidate members, only proper
motions {\it or} radial velocities are available.  Several methods
have been developed in this century to disentangle moving group stars
from a field star population using proper motion data alone. One is
the {\it convergent point method}, which exploits the perspective
effect that makes the proper motions of the group stars point towards
a convergent point on the sky. This method has been used extensively
to determine membership of moving groups (e.g., Blaauw~1946; van
Bueren 1952; Jones~1971; de Bruijne 1999). The {\it vector point
diagram method} also uses proper motions for membership selection (see
e.g., Vasilevskis, Klemola \& Preston 1958; Fresneau 1980; Jones \&
Walker 1988). Here, the fact that the common space motion of a moving
group results in similar proper motions is used to determine
membership. The proper motions will in principle stand out from the
proper motion distribution of the field stars in the vector point
diagram and thus the member stars can be separated.

These traditional methods have several shortcomings. The convergent
point method is useful when applied to a region of the sky where the
members of the group represent a significant fraction of the
sample. The presence of more than one moving group within the sample
affects the performance of this method. The vector point diagram
method is constrained to small regions of the sky, since perspective
effects, which shift the positions in the proper motion plane, are not
taken into account. This method is especially suited for member
selection in open clusters (see e.g., Prosser 1992; Tian, Zhao \& van
Leeuwen 1994). Neither method uses parallax information, even
when this is available and can be used to further constrain the
search for moving groups.

Prior to Hipparcos, coherent proper motion measurements for moving
groups which subtend large angles on the sky ($\ga$10 degrees) were
only available for the brightest stars. Using photographic plates to
cover such large areas introduced systematic errors due to the
uncertainties in combining the photographic plates and in the mostly
ill-defined plate corrections. Meridian telescopes were thus needed to
obtain reliable proper motions. However, this procedure could only be
used for stars brighter than $\sim$$7^{\rm th}$ mag because stability
criterions limited the telescope size. This has long hampered the
study of extended moving groups, in particular for the nearby OB
associations where reliable membership determination, using proper
motion data, has been made previously only for spectral types earlier
than $\sim$B5 (see e.g., Blaauw 1946; Bertiau 1958; Jones 1971).

The Hipparcos mission has vastly improved this unsatisfactory state of
affairs. The satellite was launched on 8 August 1989 and ended its
mission on 15 August 1993, obtaining accurate positions, parallaxes,
proper motions, and photometry, for $118\:218$ pre-selected
stars. Information on multiplicity and variability was obtained as
well.  A detailed description of the mission objectives and results
can be found in the Hipparcos Catalogue (ESA~1997). The small median
probable errors, $\sim$1~mas in position and parallax and
$\sim$1$~{\rm mas}~{\rm yr}^{-1}$ in proper motion, together with
negligible systematic errors, $\la$0.1~mas, make it an excellent data
base to search for moving groups and identify their members.

The high quality of the measurements in the Hipparcos Catalogue has
prompted the search for new, better methods, to identify moving
groups. Chen et al.\ (1997) introduced a non-parametric kernel
estimator to identify clustering in a 4-dimensional space of spatial
velocities and stellar ages. Powerful as this method is, it requires
the extra knowledge of radial velocities, and stellar ages, to specify
a unique location for each star in this 4-dimensional space. The
limited availability of this additional information severely restricts
the use of this method. Other methods to find moving groups or
investigate the velocity structure in the Solar neighbourhood have
been introduced by e.g., Chereul et al.\ (1998, 1999) and Dehnen
(1998). Here we present a new non--parametric method to identify
moving groups and to assess individual membership.  This method uses
the five astrometric parameters measured by Hipparcos, with no
additional information being required, and can thus be applied to the
full Hipparcos Catalogue. It applies in principle to any type of
moving group (cluster, OB association, or supercluster); e.g., this
paper presents results on the Hyades and IC2602 open clusters, while
results on the nearby OB associations can be found in de Zeeuw et al.\
(1999).

The paper is organized as follows. The next section describes the new
method for the detection of and member selection in moving groups
based on Hipparcos data. In \S \ref{sec:tests} the method is tested
using synthetic data. \S \ref{sec:hyades_ic2602} presents a realistic
test of the method on the open clusters IC2602 and the Hyades. We end
with a summary and discussion in \S \ref{sec:summary}.
%
%
\section{The Spaghetti Method}\label{sec:spa}
Our method identifies groups of stars that share kinematics within a
given velocity dispersion, and assesses the statistical significance
of such groupings. It then assigns membership for stars to the
candidate moving groups. The novel feature of the method is that it
uses {\it all} and {\it only} the astrometric parameters in the
Hipparcos database: positions, parallaxes, and proper motions. In this
regard it is qualitatively different from the classical convergent
point and vector point diagram methods that use proper motion
information only. It can also be extended to include radial velocity
information, as it becomes available. Finally, as the search is
performed in velocity space, this method does not require the members
of identified moving groups to be restricted to limited areas on the
celestial sphere.

The essence of the method is the recognition that the five
astrometric parameters --- position on the sky $(\alpha, \delta)$,
parallax $(\pi)$ and proper motion
$(\mu_{\alpha*},\mu_{\delta})$, where $\mu_{\alpha*} =
\mu_\alpha\cos\delta$ --- do not determine the position of the star in
the 6-dimensional phase-space uniquely. The parallax and the star's
position on the sky determine the three-dimensional spatial
coordinates of the star; but only two of the velocity components are
determined by the proper motion and the parallax. No information on
the third velocity component, the radial velocity, is assumed to be
available. One basic difficulty is that the two measured velocity
component directions are not the same for different stars, as each
measured pair lies on a plane orthogonal to the unique line of sight
to the corresponding star.

The five astrometric parameters thus define a line in velocity space,
orthogonal to the measured tangential velocity and parallel to the
line of sight, on which the tip of the spatial velocity vector of the
star is constrained to be. In reality, this line has a thickness set
by the uncertainties in the astrometric quantities
(Fig.~\ref{fig:spaghetti}). This cylinder in velocity space (which we
will refer to as `spaghetti') thus represents a probability
distribution for the spatial velocity of the star.

For each star we can define a spaghetti in velocity space.  Spaghettis
corresponding to stars moving with the same spatial velocity will
intersect in one point in velocity space --- the velocity of the group
--- regardless of the star's position on the sky. A moving group will
thus appear as a region in velocity space where a significant number
of spaghettis intersect. The fact that, even in the absence of real
moving groups, random intersections are expected, sets the lower limit
to the number of members in moving groups that can be detected on any
given background distribution. 

The line in velocity space, defined above, is also used by
Chereul et al.\ (1998, 1999). However, their wavelet analysis of the
velocity structure in the Solar neighbourhood requires 3-dimensional
velocity information. In order to use stars without measured radial
velocities they create discrete `artificial' velocities which lie on
this line. Furthermore, they do not take the covariance matrix of the
Hipparcos data into account, i.e., they ignore the thickness and shape
of the spaghetti.

\subsection{From Hipparcos to cylinders in velocity space}\label{sec:cyl}
\begin{figure}
  \centerline{ \psfig{file=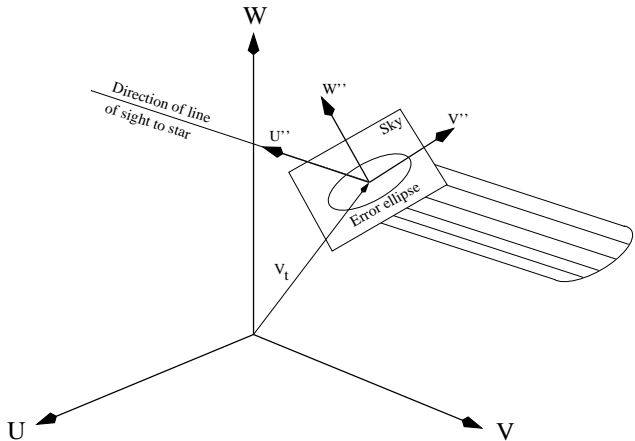,width=8.35cm}}
  \caption{Cylinder in velocity space $(U,V,W)$ determined by the
  Hipparcos data. The tangential velocity $(\bmath{v}_t)$ determines
  the offset from the origin of the central axis of the cylinder. The
  orientation of the cylinder is determined by the direction of the
  line of sight. The errors and correlations in the tangential
  velocity result in an elliptical cross section of the cylinder,
  which is infinitely long, as no kinematical information in the
  radial direction is assumed to exist. $(U,V,W)$ can be any set of
  orthogonal axes in velocity space. It is customary to choose $U$ in
  the direction of the Galactic centre, $V$ in the direction of
  Galactic rotation, and $W$ towards the north Galactic pole. The
  coordinate system $(U'',V'',W'')$ is described in the text.}
  \label{fig:spaghetti}
\end{figure} 
The direction of the spaghetti axis in velocity space is determined by
the position $(\alpha, \delta)$ of a star on the sky. This is the
direction in which we lack velocity information. The parallax $(\pi)$
and the proper motion $(\bmath{\mu} = [\mu_{\alpha^*},\mu_{\delta}])$
determine the tangential velocity in ${\rm km}~{\rm s}^{-1}$,
$\bmath{v}_t = 4.74 \bmath{\mu} / \pi$ where $\bmath{\mu}$ is in ${\rm
mas}~{\rm yr}^{-1}$ and $\pi$ is in mas. The tangential velocity sets
the offset of the central axis of the spaghetti from the origin in
velocity space (see Fig.~\ref{fig:spaghetti}). The axis $\bmath{s}$ is given by
\begin{equation}
{\bmath{s}} = {\bmath{v}_t} + \lambda {\bbeta},
\label{eq:def_spa}
\end{equation}
where the direction of the line is given by $\bbeta$, the unit
vector in the radial velocity direction, and $\lambda$ is a free scalar
parameter. When $\lambda$ is equal to the radial velocity, $\bmath{s}$
equals the star's spatial velocity. We use an orthogonal velocity
coordinate system $(U,V,W)$, with the Solar-system barycentre at the
origin, $U$ directed towards the Galactic centre, $V$ in the direction
of Galactic rotation, and $W$ towards the north Galactic pole.

The axis $\bmath{s}$ becomes a cylinder in velocity space with an
elliptical cross section due to the errors on the tangential
velocity. The full covariance matrix of the astrometric parameters is
taken into account in the determination of these errors (see ESA 1997,
Vol.\ 1 \S 1.5.6). The semi-major axis, axis-ratio, and orientation of
the ellipse can be calculated using the covariance matrix ${\bf C}$ of
the tangential velocity\footnote{We use standard
error propagation to obtain the probability distribution of the
tangential velocity. However, for stars with unreliable parallaxes,
$\pi/\sigma_{\pi} \la 0.5$, this is not the best representation. In
principle the following integral, which can be evaluated analytically,
should be calculated:
\[
 P_{\bmath{v}_t} = \int\limits_{-\infty}^{\infty} {\rm d} \tilde{\pi}
                  \frac{\tilde{\pi}^2}{\kappa^3} \frac{1}{\sqrt{(2\pi)^3 |\bf C|}} 
                  \exp[-1/2 (\bmath{\xi} - \bmath{\xi}_o)^T {\bf C}^{-1} 
                            (\bmath{\xi} - \bmath{\xi}_o)],
\]
with $\kappa = 4.74...$, $\bmath{\xi} =
(\tilde{\pi},\mu_{\alpha\ast},\mu_\delta) = (\tilde{\pi},\tilde{\pi}
v_\alpha / \kappa,\tilde{\pi} v_\delta / \kappa)$, $\bmath{\xi}_o$ the
observables, ${\bf C}$ its covariance matrix, and $\tilde{\pi}$
denotes the parallax in order to avoid confusion with $\pi = 3.14...$. This
results into a skewed probability distribution of the tangential
velocity, and a mean tangential velocity which differs from $\bmath{v}
= \kappa \bmath{\mu}/\tilde{\pi}$. We did not implement this representation of
the tangential velocity error because the difference with the standard
approach is only noticeable for $\tilde{\pi} \la 2$~mas, for a typical
Hipparcos measurement. In this regime the parallax, and thus the
tangential velocity, is only accurate to 50\% or less and consequently
``useless'' for the Spaghetti method. Furthermore, any set of
physically related stars which still have a common space motion will,
most likely, occupy a region of configuration space less than 100~pc
in diameter. Differential Galactic rotation will destroy the coherent
velocity structure on larger scales. The change in skewness and mean
of the above approach does not change notably over such scales. We
thus expect that the standard error propagation is not hampered by any
systematic shifts in tangential velocity with distance.}.

For simplicity, we introduce a new coordinate system $(U'',V'',W'')$
such that the $U''$-axis coincides with the axis of the cylinder
($\bmath{s}$), and the $V''$- and $W''$-axis correspond to the major
and minor axes, respectively, of the elliptical cross section (see
Fig.~\ref{fig:spaghetti}). The surface equation for the cylinder is
simplest in this coordinate frame and allows us to define a
probability function. The coordinate transformation is done in two
parts. First, we apply a rotation, ${\bf T}$, and translation to an
intermediate coordinate system $(U',V',W')$. The $U'$-axis coincides
with the axis of the cylinder, while the $V'$- and $W'$-axis are not
yet aligned with the major and minor axis, respectively. We then
translate the origin of the coordinate system to coincide with the tip
of the tangential velocity vector. The translation has no effect on
the covariance matrix, while the rotation affects ${\bf C}$ as follows
\begin{equation}
{\bf \hat{C}} = {\bf T} \: {\bf C} \: {\bf T}^{\rm T},  
\label{eq:c_hat}
\end{equation}
where ${\bf \hat{C}}$ is the covariance matrix in the intermediate
coordinate system $(U',V',W')$.  ${\bf T}^{\rm T}$ is the transpose of ${\bf
T}$. Second, we apply a rotation around the $U'$-axis over an angle
$\eta$, ${\bf T}_{U'}(\eta)$, which can be obtained from the
sub-matrix ${\bf \hat{C}}'$ consisting of the $V'$ and $W'$ velocity
components. This matrix determines the shape of the elliptical surface
of the cylinder. The semi-major axis, $a$, axis ratio, $q$, and
rotation angle with respect to the $V'$-axis, $\eta$, of the ellipse
can be written as
\begin{eqnarray}
{\bf \hat{C}}' & = & \left[ \begin{array}{cc}
		     \sigma_{V'}^2 & \rho \: \sigma_{V'} \sigma_{W'} \\
		     \rho \: \sigma_{V'} \sigma_{W'} & \sigma_{W'}^2 \\
		     \end{array} \right ], \nonumber\\
a^2            & = & 
		     \frac{2(1-\rho^2)\sigma_{V'}^2\sigma_{W'}^2}
		     {\sigma_{V'}^2+\sigma_{W'}^2 - 
		     \sqrt{(\sigma_{V'}^2+\sigma_{W'}^2)^2
		     -4(1-\rho^2)\sigma_{V'}^2\sigma_{W'}^2 }}, \nonumber\\
q^2            & = & 
		     \frac{a^4}{(1-\rho^2)\sigma_{V'}^2\sigma_{W'}^2}, \nonumber\\
\eta           & = & \frac{1}{2} \arctan \left ( 
		     2 \rho \frac{\sigma_{V'}
		     \sigma_{W'}}{\sigma_{V'}^2-\sigma_{W'}^2} \right ),
\label{eq:c_hat_prime}
\end{eqnarray}
where $\rho \equiv \rho_{V'}^{W'}$.  After the rotation we obtain the
$(U'',V'',W'')$ coordinate system in which the $V''$- and $W''$-axis
are aligned with the major and minor axes of the error ellipse,
respectively. The covariance matrix in the final system can be
calculated with eq.~(\ref{eq:c_hat}), using ${\bf T}' = {\bf
T}_{U'}(\eta) \: {\bf T}$ instead of ${\bf T}$. The positional
accuracy in the Hipparcos Catalogue results in an error in the
direction of the cylinder $(\bbeta)$ on the order of $10^{-8}$. We
neglect this error.

\subsection{The spaghetti density in velocity space}\label{sec:sdf}
To find the points in velocity space where the number of intersecting
cylinders are highest, we define the {\it Spaghetti Density Function} in
velocity space $(U,V,W)$ as,
\begin{equation} 
{\rm SDF}(U,V,W) = \sum\limits_{i = 1}^{N} \frac{1}{2 \pi a_i'^2 q_i'} 
   \exp \left[ -\frac{1}{2}\frac{V''^2 + W''^2/q_i'^2}{a_i'^2} \right].  
\label{eq:sdf}
\end{equation} 
Here $V''$ and $W''$ are the velocity components in the $(U'',V'',W'')$
system; $a'_i$ and $q'_i$ are the semi-major axis and axis ratio of
the error ellipse as defined below, respectively, and $N$ is the number
of stars. Each term in the summation is interpreted as the probability
function for the spatial velocity vector of the corresponding star,
and it contains all the kinematic information that the Hipparcos
Catalogue provides for that star. The SDF is normalized such that
for each cylinder $i$, the integral of the $i^{\rm th}$ component of
the SDF over a plane perpendicular to its axis is unity.

The elliptical spaghetti cross section is constrained to have a
minimum size, $\sigma_{\rm int}$, thus $a_i' = {\rm MAX}(\sigma_{\rm
int},a_i)$, $b_i' = {\rm MAX}(\sigma_{\rm int},b_i)$ and $q_i' =
b_i'/a_i'$.  The reasons for this thickening of the thinnest
spaghettis are threefold. First, moving groups have intrinsic velocity
dispersions ranging from a few tenths of ${\rm km~s}^{-1}$ in open
clusters (e.g., Dravins et al.\ 1997; Perryman et al.\ 1998) to a few
${\rm km~s}^{-1}$ in OB associations (see e.g., Jones \& Herbig 1979;
Hartmann et al.\ 1986; Mathieu 1986; Tian et al.\ 1996). When the mean
spaghetti thickness is less than the intrinsic velocity dispersion in
a moving group, the typical separation between the axes of the
spaghettis is larger than the mean spaghetti thickness. As a result
the moving group will not generate a peak in the SDF. This can be
solved by artificially increasing the thickness of the spaghettis.
Second, the errors on the tangential velocity components are highly
correlated due to the way the parallax enters in the conversion of
proper motion to velocity. This causes the spaghettis to be very thin
in one direction, and depending on the orientation of the spaghettis,
the moving group may not generate a peak in the SDF. Third, long
period binaries artificially increase the velocity dispersion of a
moving group. For binaries with periods longer than the four year
observing period of Hipparcos, only part of the orbit was observed and
the proper motion may thus be different from that of the centre of
mass of the binary system (see Wielen et al.\ 1997). This effect is
especially noticeable for massive equal mass binaries within 200~pc of
the Sun and increases the velocity dispersion. 

The procedure then is to search for maxima, or peaks, in the SDF. We
evaluate the SDF on a Cartesian grid of $4~{\rm km}~{\rm s}^{-1}$
spacing, and start a steepest gradient search from every grid point
which is a local maximum to find the $(U,V,W)$ positions of
peaks. This sampling is dense enough to find every peak in the SDF as
it is of the same order as the structure in the SDF, whose scale is
set by the typical errors in tangential velocity and the internal
dispersion, $\sigma_{\rm int}$.

\subsection{Peak significance}\label{sec:peaks}
In order to assign a statistical significance to peaks in the
SDF obtained from the data, it is necessary to build the
`null hypothesis' and obtain its expected distribution of peaks.
The ideal null hypothesis is the Hipparcos database that would be
obtained when observing a Galaxy in which no moving groups exist.

Ideally, we would do a series of Monte Carlo experiments in which
several random realizations of the Hipparcos Input Catalogue are
generated from the null model.  The resulting observed properties
would be reduced with the techniques used for the real data to derive
the final astrometric quantities and their associated uncertainties
and correlations. Such a procedure is enormously complex, and for
this, a suitable approximate null hypothesis must be found. Therefore,
we decided to generate random realizations of the proper motion
components alone for the stars in the real data set, leaving all other
quantities fixed. We use the Schwarzschild ellipsoid as the local
velocity distribution for our null model of the Galaxy without moving
groups. The first and second moments, and the vertex deviation were
taken from an analysis of local Hipparcos main-sequence stars by
Dehnen \& Binney (1998). This model also includes their values of the
Solar motion and asymmetric drift. The Oort constants for the Galactic
rotation are taken from Feast \& Whitelock (1997). This procedure
should produce a reasonable approximation, since it is only the
information that bears the signature for grouping in velocity space,
the proper motion information, that has been obtained from the null
model, leaving everything else intact. The kinematic model is based on
an unbiased sample of main-sequence stars within 100~pc of the
Sun. Stars at larger distances, giants, and super giants may have
different kinematic properties and using those of the nearby
main-sequence stars in the simulations results in some systematic
effects in the null hypothesis. For more details on this procedure see
\S 3.4 in de Zeeuw et al.\ (1999).

The expected distribution of peak heights is obtained from 100 Monte
Carlo experiments and the significant peaks in the real data are
obtained by direct comparison. An additional difficulty comes from the
fact that the variation in density of stars in velocity space produces
a changing background in the SDF, thus making the statistical
significance of a peak dependent on position in velocity space. To
address this problem, we obtain the median SDF from the set of all
Monte Carlo realizations of the null model sampled on a grid of 4
${\rm km~s}^{-1}$ spacing. The median SDF is subtracted from both the
SDF of the real data and of the experiments.

\subsection{Membership}\label{sec:membership}
Having located a significant peak in the SDF we determine membership
for all stars. The fraction of the spaghetti which lies within a
sphere of radius $R_s$ in velocity space centred on the peak is taken
as a measure of membership. This fraction is evaluated using the volume
integral of the spaghetti over the sphere. We take
\begin{equation}
R_s = \sqrt{\sigma_{\rm int}^2 + \sigma_{\rm median}^2},  
\label{eq:rs}
\end{equation} 
so that $R_s$ is a measure of the extent of the peak resulting from
the median error on the tangential velocity, $\sigma_{\rm median}$,
and the estimated internal velocity dispersion, $\sigma_{\rm
int}$. $\sigma_{\rm median}$ is the typical thickness of a spaghetti
and is set to the median of the spaghetti semi-major axes for which
the SDF is calculated. $\sigma_{\rm median}$ can also be calculated
for stars in a certain distance range if some {\it a priori} knowledge
on the moving group distance is available, and can be improved upon by
iteration. The integral over the $U''$-axis of the volume integral can
be done analytically. This leaves the following double integral
\begin{eqnarray}
S & = & \frac{1}{2 R_s} \,\int\!\!\!\int_{C} 
	2 \sqrt{R_s^2 - (V'' - V''_0)^2 - (W'' - W''_0)^2} \times \nonumber\\
  &   & \quad \quad \frac{1}{2\pi a^2 q}  
    \,\exp \left [ -\frac{1}{2}\frac{V''^2 + W''^2/q^2}{a^2} \right ]
    dW''dV'',
\label{eq:s}
\end{eqnarray}
where $a$ is the semi-major axis of the spaghetti and $q$ the axis
ratio. Here $C$ is the surface enclosed by a circle centred on
$(V''_0,W''_0)$ with radius $R_s$. $(U''_0,V''_0,W''_0)$ are the
coordinates of the peak. As a spaghetti is normalized to unity in the
plane perpendicular to its axis, we divided the integral by $2 R_s$
for $S$ to range between 0 to 1.

$S$ can be interpreted as a conditional probability. However, it is
not straightforward to compare conditional probabilities, wich makes
them difficult to use for membership assignment. Therefore we
consider as members only those stars for which $S$ exceeds a threshold
value $S > S_{\rm min}$ (see discussion in \S
\ref{sec:results_membership}). $S$ depends on distance; for two
spaghettis with exactly the same central axis but different thickness,
$S$ will be smallest for the thickest spaghetti. Since the thickness
of the spaghettis increases with increasing distance --- the errors on
the tangential velocity increase --- $S$ will decrease and thus
depends on distance. Any selection method which includes the parallax
suffers from this effect.

\section{Tests on synthetic data}\label{sec:tests}
We now describe tests of the Spaghetti method on synthetic data
consisting of a field star population and one moving group, and one
data set containing field stars and two moving groups.

\subsection{Synthetic data sets}\label{sec:synth_data}
We use the procedure described in \S \ref{sec:peaks} to generate a
sample of field stars. We draw a star from the Hipparcos Catalogue
and, if the star lies in the requested field (see
Table~\ref{tab:simulations}), replace its observed proper motion
with a proper motion consistent with our kinematic model for the Solar
neighbourhood (\S \ref{sec:peaks}). Again, all other information,
e.g., position, parallax, errors, and correlations, are not altered.

We generate 20 random realisations for each of 24 different setups in
which the field centre and distance of the cluster are varied. For
each distance we adopt a different field size and different numbers of
field and cluster stars (see Table~\ref{tab:simulations}). The total
number of stars is chosen such that there are about 3 stars per square
degree, as in the Hipparcos Catalogue. Distant clusters subtend
smaller angles on the sky, allowing a smaller field, and as only the
brightest members are visible the number of cluster stars decreases
with distance. The adopted numbers are consistent with those found by
de Zeeuw et al.\ (1999) for the nearby OB associations. Varying the
Galactic longitude of the field centre allows us to investigate the
effect of Solar motion and Galactic rotation. The positions of the
moving group members are drawn from a sphere of constant density with
a radius of 15~pc, which is a typical size for a moving group (see
e.g., Blaauw 1991; Perryman et al.\ 1998). The velocity of each star
consists of four components: Solar motion, Galactic rotation,
streaming velocity, and an internal velocity dispersion. The streaming
velocity is defined as the spatial velocity of a moving group with
respect to its own standard of rest. The Solar motion and Galactic
rotation are the same as those used for the field stars. The streaming
velocity, which is different in each data set, is drawn uniformly from
a sphere of 10~${\rm km}~{\rm s}^{-1}$ radius.  This is a reasonable
representation of the velocity distribution of molecular clouds, which
have a typical dispersion of 6--8~km~s$^{-1}$ (see e.g., Dickey \&
Lockman 1990; Burton, Elmegreen \& Genzel 1992). The internal velocity
dispersion is represented as a Gaussian with $2~{\rm km}~{\rm s}^{-1}$
standard deviation in each coordinate. Position and velocity are then
converted into position on the sky, parallax, and proper motion. We
then perturb the parallax and proper motion with errors drawn from a
Gaussian error distribution with 1~mas and 1~${\rm mas}~{\rm yr}^{-1}$
standard error, respectively.

\begin{table}
  \caption{The 24 different setups of the synthetic data sets used to
  test the Spaghetti method. For every setup 20 random realizations
  are created resulting in a sample of 480 data sets. Four distances
  are used for each of the following 6 field centres: $\ell = 0^\circ\!,
  60^\circ\!, 120^\circ\!, 180^\circ\!, 240^\circ\!, 300^\circ\!$ and
  $b = 0^\circ\!$.  The other characteristics of each data set, such
  as field size and the number of field and cluster stars, depend on
  the distance. The total number of stars per field is consistent with
  3 stars per square degree, as in the Hipparcos Catalogue. The size
  of the simulated cluster is 15~pc in radius for each data set.}
\begin{center}
\begin{tabular}{cccc} 
\hline Distance & Field stars &
  Cluster stars & Field size \\ pc & \# & \# & \\ [1.0ex] 150 & 1000 &
  200 & $20^\circ \times 20^\circ$ \\ 300 & \phantom{1}575 & 100 &
  $15^\circ \times 15^\circ$ \\ 450 & \phantom{1}250 & \phantom{1}50 &
  $10^\circ \times 10^\circ$ \\ 600 & \phantom{1}275 & \phantom{1}25 &
  $10^\circ \times 10^\circ$ \\ \hline
\end{tabular}
\end{center}
\label{tab:simulations}
\end{table}
\begin{figure*}
  \centerline{\psfig{file=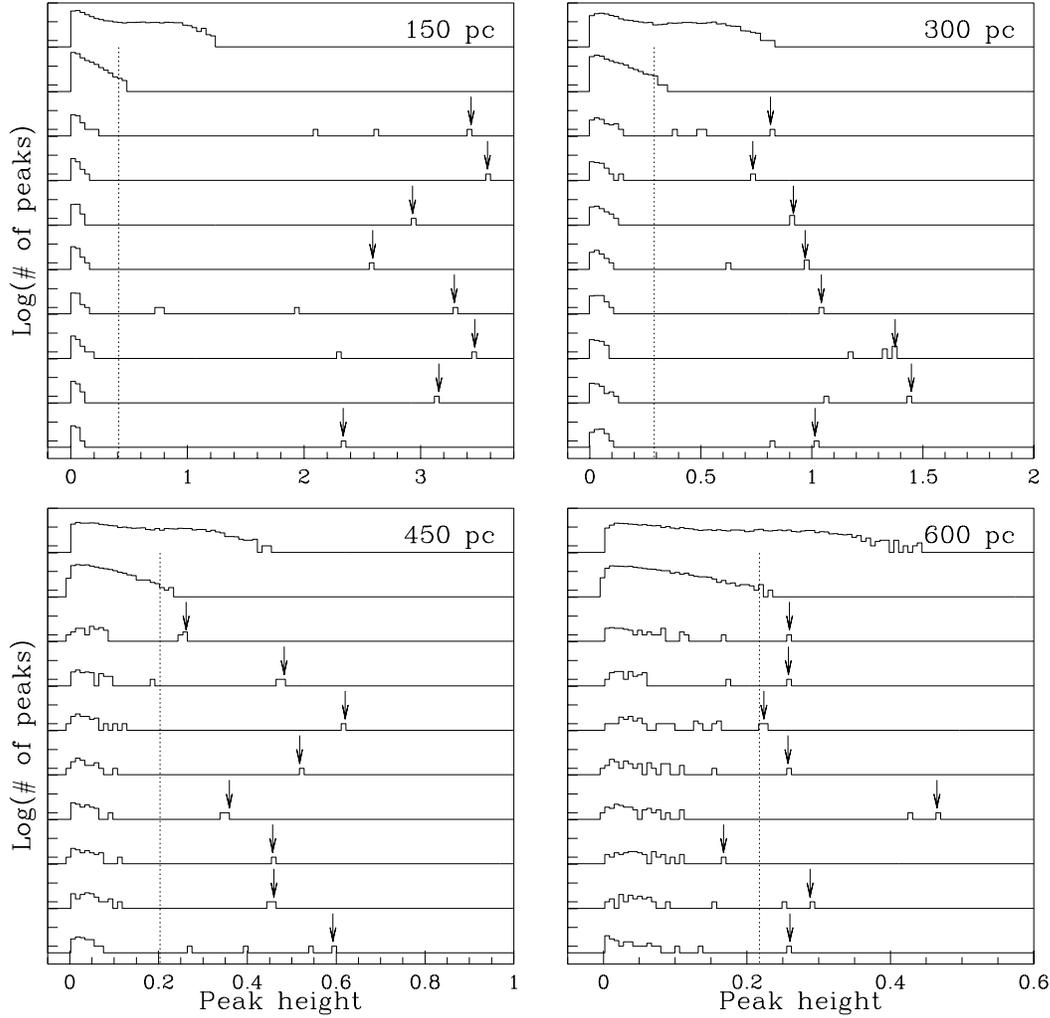,width=14cm}}
  \caption{The four panels show the peak height distribution for 8
  synthetic data sets for the $(\ell,b) = (0^\circ\!,0^\circ\!)$ field
  at 4 different distances each. The top histogram in each panel shows
  the distribution of peak heights in the 100 Monte Carlo realizations
  for the null hypothesis, no moving groups, of the synthetic data
  sets. The ticks on the ordinate have a spacing of two decades, at 1
  and 100 peaks. The second histogram is the peak height distribution
  for the null hypothesis after subtraction of the SDF background. The
  following 8 histograms give the peak height distribution, after the
  SDF background has been subtracted, for the first 8 of the 20
  synthetic data sets of these particular setups. The peak
  corresponding to the moving group is indicated by an arrow. The
  dotted line indicates a 99.9 per cent chance of finding a smaller
  peak height. See text for further explanation.}
\label{fig:peaks}
\end{figure*}

\subsection{Results}\label{sec:results}

\subsubsection{Peak significance}\label{sec:results_peaks}

The distributions of peak heights in the synthetic data sets for the
direction $(\ell,b) = (0^\circ\!,0^\circ\!)$, and distances of $150$,
$300$, $450$, and $600$~pc are shown in Fig.~\ref{fig:peaks}. The top
histogram in each panel shows the combined peak height distribution
for the 100 Monte Carlo simulations of the null hypothesis before
subtraction of the SDF background. Although most peaks have small peak
heights there is a plateau at larger values. After subtracting the SDF
background, the peak height distribution, shown in the second
histogram from the top, is nearly exponential. The Monte Carlo peak
height distribution is the result of chance intersections of
spaghettis in velocity space; extending it to higher peak values would
require an order of magnitude more simulations. With current
workstations this is not practical. This makes it difficult to assign
a realistic significance value to peaks higher than 0.5. The dotted
lines in Fig.~\ref{fig:peaks} indicate the 0.1 per cent chance for a
peak of at least that peak height to be generated in a data sample of field
stars only.

The following 8 histograms in each panel show the peak hight
distribution, corrected for the SDF background, for the first 8 of the
20 realisations of these particular situations. In each panel the peak
distribution follows the null hypothesis peak distribution for the
small peak values, $\la 0.1$. However, the panels also show one or
more peaks around 3.0, 1.0, 0.4, and 0.2 for 150, 300, 450, and
600~pc, respectively. In all cases one of these peaks corresponds to
the moving group, indicated by the arrows. These peaks stand well
clear from the Monte Carlo peak distribution. The height of the moving
group peak decreases with distance. More distant clusters are less
significant (\S \ref{sec:sdf}).  Furthermore, the peak height also
depends on the number of moving group stars present in the data set,
the velocity dispersion, and the data quality.

The Spaghetti method sometimes finds more than one significant peak in
a synthetic data set containing only one moving group. This is due to
the parallelism of the spaghettis --- as we are looking at a specific
field on the sky all spaghettis have similar directions. This causes
the moving group peak in the SDF to be stretched in the radial
direction (see also Fig.~\ref{fig:hyades} panel [j]). Small random peaks
generated by the field star population superposed on this
larger peak will be classified as significant based on peak
height only. In general, the highest of this set of peaks is centred
on the moving group velocity, while the significant artificial peaks
are all aligned with the line of sight direction passing through the
moving group velocity. This distinguishes artifical peaks from multiple
moving groups.

\subsubsection{Membership threshold and success rate}\label{sec:results_membership}

\begin{figure}
  \psfig{file=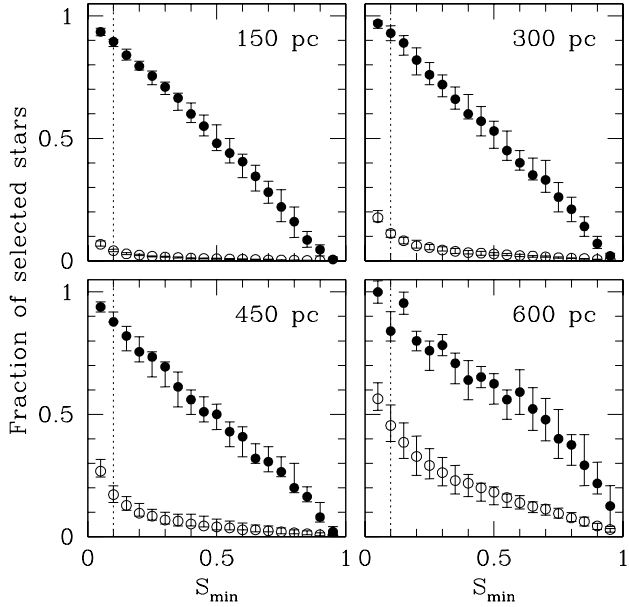,width=8.35truecm}
  \caption{Fraction of moving group, closed circles, and field, open
  circles, stars selected by the Spaghetti method as a function of
  membership threshold $S_{\rm min}$. These results are for the field
  $(\ell,b) = (0^\circ\!,0^\circ\!)$ and $\sigma_{\rm int} = 2~{\rm
  km~s}^{-1}$. The distance of the cluster is 150,
  300, 450, and 600~pc with the corresponding properties as described
  in Table~\ref{tab:simulations}. The figure displays the median of 20 random
  realizations and the errors are estimated by the interquartile
  range.}
\label{fig:s_min}
\end{figure}

Fig.~\ref{fig:s_min} shows, for one of the fields, $(\ell,b) =
(0^\circ\!,0^\circ\!)$, the fraction of selected field and cluster
stars versus $S_{\rm min}$ (see \S \ref{sec:membership}). We used the
stars within a distance range of 200~pc centred on the cluster
distance to determine $\sigma_{\rm median}$ and hence $R_s$ (see \S
\ref{sec:membership}). We used $\sigma_{\rm int} = 2~{\rm
km~s}^{-1}$. A sharp increase can be seen in the fraction of selected
field stars at $S_{\rm min} \approx 0.1$. On the other hand, the
number of selected cluster stars rises steadily for decreasing $S_{\rm
min}$. This trend led us to pick $S_{\rm min} = 0.1$ as the
border-line between member and non-member. It is clear that the value
of $S_{\rm min}$ can be adjusted for specific problems. The results do
not depend sensitively on position on the sky.

\begin{figure}
  \psfig{file=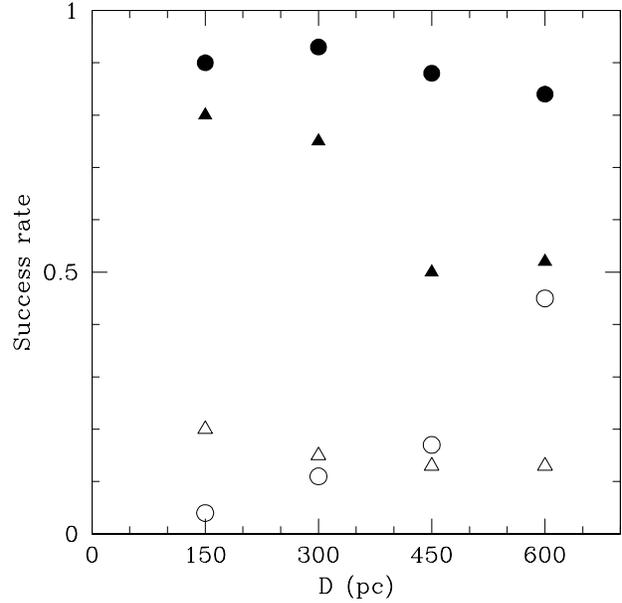,width=8.35truecm}
  \caption{Success rate of the Spaghetti method compared to the
  classical convergent point method (by de Bruijne 1999) as a function
  of distance. The filled circles and triangles represent the fraction
  of cluster stars found in the Monte Carlo simulations by the
  Spaghetti and convergent point method, respectively. The open
  circles and triangle represent the fraction of field stars found
  for both methods.}
\label{fig:spa_det}
\end{figure}

The threshold value $S_{\rm min} = 0.1$ leads to the selection of
90/4, 93/11, 88/17, and 84/45 per cent of the cluster/field stars for
150, 300, 450, and 600~pc respectively (see
Fig.~\ref{fig:spa_det}). These success rates can be compared with
those obtained by the refurbished convergent point method of de
Bruijne (1999) using similar simulations. He finds 80/20, 75/15,
50/13, and 52/13 per cent of the cluster/field stars for the same
distances. The percentage of selected cluster stars is higher for the
Spaghetti method and remains roughly constant over the sampled
distance range. The percentage of selected field stars in the
Spaghetti method increases with distance, especially for the most
distant data sets, whereas in the convergent point method this
percentage remains constant. Only for the most distant data sets does
the Spaghetti method select many more field stars than the convergent
point method. However, the number of selected cluster stars remains
above 80 per cent. Although the contamination by field stars increases
with distance, the Spaghetti method samples a larger fraction of the
cluster than the convergent point method. The increase in selected
field stars with distance is due to the increasing tangential velocity
errors, corresponding to an increase in the thickness of the
spaghettis. This results in a larger value of $R_s$, the radius of the
sphere in velocity space used for the membership determination (see \S
\ref{sec:membership}). The fraction of selected cluster stars shows
the same trend with $S_{\rm min}$ at any distance because the typical
thickness of these spaghettis determines $R_s$, while the number of
selected field stars increases because $R_s$ is not related to those
stars. Field stars are located at all distances and their spaghettis
span a whole range of thicknesses. Especially the nearest field stars,
having thin spaghettis, will be selected more easily if $R_s$ is
large. This also puts a limit, $\sim$750~pc, on the distance at which
the Spaghetti method can be used reliably using the Hipparcos
data. The typical error on the tangential velocity at 750~pc,
$\sim$$20~{\rm km}~{\rm s}^{-1}$, is of the same order as the large
scale structure in velocity space. Other diagnostics, e.g., the
parallax distribution or photometric information of the selected stars
or a combination with another selection method (see e.g., de~Zeeuw et
al.\ 1999), can be used to lower the number of selected field stars.

\begin{figure}
  \centerline{\psfig{file=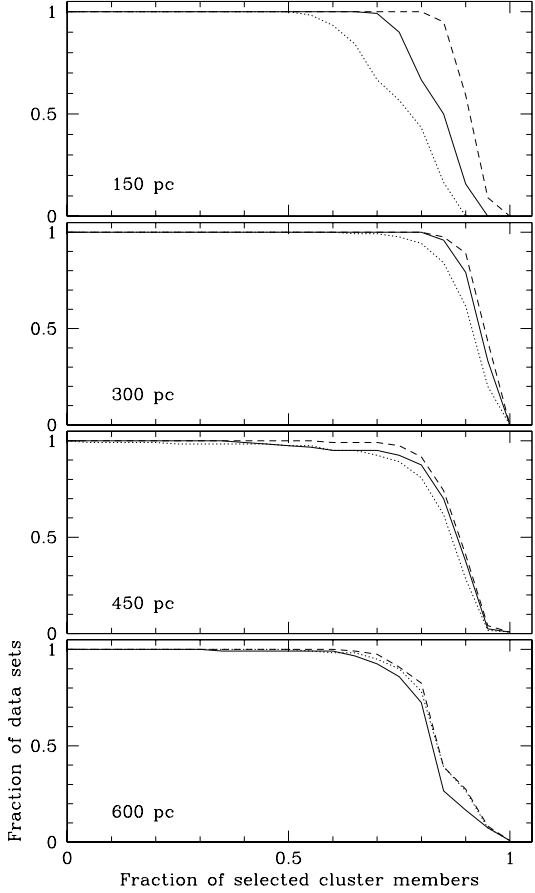,height=12cm}}
  \caption{
  The panels show the success rate of the Spaghetti method as a
  function of the percentage of selected cluster members, using
  $S_{\rm min} = 0.1$ (i.e.\ for 150~pc and $\sigma_{\rm int} = 2~{\rm
  km~s}^{-1}$ we find 50 per cent of the cluster members in all data
  sets and 85 per cent in half of the data sets). The panels are
  divided according to distance. The solid lines indicate the results
  of the member selection for $\sigma_{\rm int} = 2$~km~s$^{-1}$,
  equal to the internal velocity dispersion in the simulations. The
  dotted and dashed lines indicate the results for $\sigma_{\rm int} =
  1$ and 3~km~s$^{-1}$, respectively. The lower three panels show that
  for these distances $\sigma_{\rm int}$ is not important in the
  member selection; the errors on the tangential velocity are larger
  than the velocity dispersion. For moving groups at $\sim$150~pc,
  $\sigma_{\rm int}$ is important and the panel shows that its value
  should be taken somewhat larger than the internal velocity dispersion.}
\label{fig:success}
\end{figure}

Fig.~\ref{fig:success} shows the success rate of the Spaghetti method
as a function of distance, the value of $\sigma_{\rm int}$, and the
percentage of selected cluster members. It shows in how many of the
data sets we select a certain percentage of cluster stars, e.g., at a
distance of 150~pc and $\sigma_{\rm int} = 2~{\rm km~s}^{-1}$ we find
85 per cent of the cluster members in 50 per cent of the data sets.
It shows that the value for $\sigma_{\rm int}$ is not important in the
member selection for moving groups beyond $\sim$300~pc. This is
expected because, at these distances, the errors on the tangential
velocity dominates $R_s$ (eq.~\ref{eq:rs}). For moving groups at smaller
distances it is important to know the internal velocity dispersion.
The top panel in Fig.~\ref{fig:success} shows that if $\sigma_{\rm
int}$ is equal to e.g., 1.5 times the velocity dispersion in the
simulated data sets the success rate is much better than when
$\sigma_{\rm int}$ is equal to this velocity dispersion. As it is
difficult to obtain reliable estimates for the internal velocity
dispersions in moving groups it is best to use a somewhat large value
for $\sigma_{\rm int}$ for the nearest moving
groups. Fig.~\ref{fig:success} also shows that the success rate
decreases with distance. This is due to the increasing errors on the
tangential velocities which decrease the `resolving power' of the
Spaghetti method at larger distance. The Spaghetti method will never
find all moving group members because there will always be stars in
the wings of the velocity distribution that fall outside $R_s$. This
argument applies to any selection process for moving groups, and
results in a characteristic number of cluster stars we find in a
typical data set of: $\sim$90 per cent for 150 to 300~pc and $\sim$85
and $\sim$80 per cent for 450 and 600~pc, respectively.

\begin{figure}
  \psfig{file=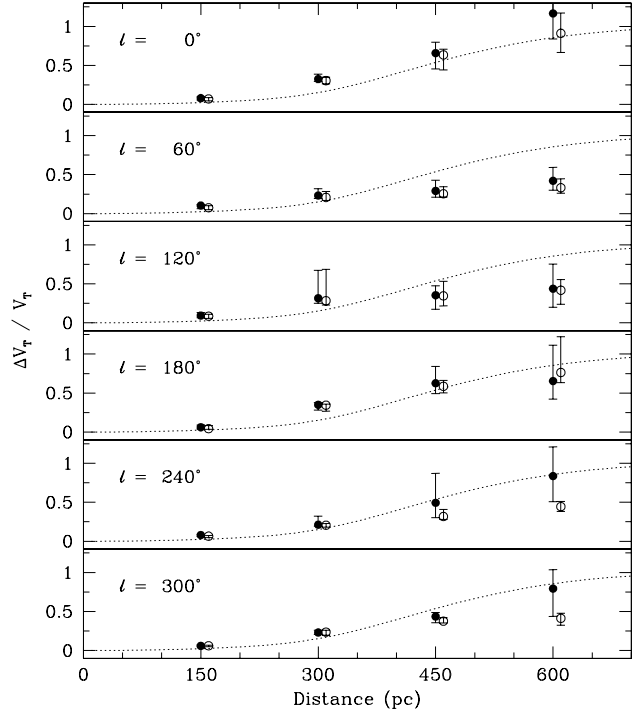,width=8.35cm} 
  \caption{The relative difference in tangential velocity $\Delta V_T
  / V_T$ of the synthetic moving cluster and the tangential velocity
  found by the Spaghetti method, as a function of distance. The filled
  dots and error bars are the median and interquartiles respectively
  for 20 data sets. The open dots, shifted by 10~pc, show the bias in
  $\Delta V_T / V_T$ as present in the Monte Carlo simulations of only
  cluster stars, so no field contamination is present. The 6 panels
  show the results for the line of sight indicated in each panel. The
  dotted lines indicate the predicted bias in tangential velocity
  resulting from the use of the parallax in its calculation.}
\label{fig:velocity}
\end{figure}

\subsubsection{Velocity}\label{sec:results_velocity}

Apart from a list of members and non-members, the Spaghetti method
produces the $(U,V,W)$ velocity of the moving group --- the velocity of
the peak in the SDF. The predicted radial velocity will have the
largest uncertainty. When a moving group is confined to a small region
on the sky, it is difficult to find the precise radial velocity due to
the parallelism of the spaghettis which stretch the peak in the
radial direction (see e.g., Fig.~\ref{fig:hyades} panel [j]). If
however, a moving group covers a large fraction of the sky the
spaghettis contributing to the peak come from a large range of
directions, and the predicted radial velocity will be much better
constrained (see e.g., the Cas--Tau association covering more than
100$^\circ\!$ by 60$^\circ\!$ on the sky [\S 7.2 in de Zeeuw et al.\
1999]).

Fig.~\ref{fig:velocity} shows the relative difference in the
tangential velocity between that of the moving group and the one found
by the Spaghetti method, $\Delta V_T/V_T = (V_{\rm spaghetti} -
V_T)/V_T$. Here $V_T$ is the tangential part of the total moving group
velocity used in the simulations, representing the group's streaming
motion, the Solar motion, and Galactic rotation. $V_{\rm spaghetti}$
is the tangential velocity found by the Spaghetti method. For every
direction, $\Delta V_T/V_T$ increases with distance. At 600~pc in the
direction $\ell = 300^\circ\!$ the offset is as large as 20~${\rm
km}~{\rm s}^{-1}$. This trend in $\Delta V_T/V_T$ is due to a bias
introduced by the parallax in the calculation of the tangential
velocity, and is similar to the bias in distance when calculated as
$1/\pi$, see Smith \& Eichhorn (1996) and Brown et al.\ (1997). To
estimate the magnitude of the tangential velocity bias we redid the
Monte Carlo simulations of the synthetic data sets without the field
stars (indicated in Fig.~\ref{fig:velocity} by the open dots). The
tangential velocity differences we find in these simulations includes
the true bias but excludes any effect introduced by the field star
population. We also calculated the proper motion due to Solar motion
and Galactic rotation in the six different directions as a function of
distance. Combined with the biased distances, as described by Brown et
al.\ (1997), this gives another measure of the bias in tangential
velocity, indicated by the dotted line in Fig.~\ref{fig:velocity} which
follows the points rather well. The most distant points in the
directions $\ell = 60^\circ\!$ and $120^\circ\!$ do not follow the
trend due to the bias in parallax only. This could be due to the
combined effects of Solar motion and Galactic rotation which results
in very small, almost zero, tangential velocity for these directions
and distances. The effects due to the selection method probably
dominate in these cases.

The spread in $\Delta V_T/V_T$ we find is of the same magnitude as the
typical error in the tangential velocity, and indicated in
Fig.~\ref{fig:velocity} by the error bars. The spread in radial
velocities is much larger and in some cases differs by as much as
100~${\rm km}~{\rm s}^{-1}$ from the real radial velocity.

Another important characteristic of the Spaghetti method is that a
simple linear expansion or contraction of a moving group will
influence the radial velocity found by our method. This is inherent to
the fact that we use proper motions and parallaxes only. Without
accurate radial velocities ($\sigma_{v_{\rm radial}} < {\rm
a~few}~{\rm km}~{\rm s}^{-1}$) it is impossible to distinguish an
expanding moving group which is stationary with respect to the
observer from a moving group which is moving towards the observer ---
the so-called perspective effect --- (Blaauw 1964). Thus, when using
the predicted radial velocity of a moving group it is important to
keep in mind that part of the radial velocity might be due to
expansion or contraction.

\subsection{Multiple groups}\label{sec:multiple_groups}

\begin{figure}
  \psfig{file=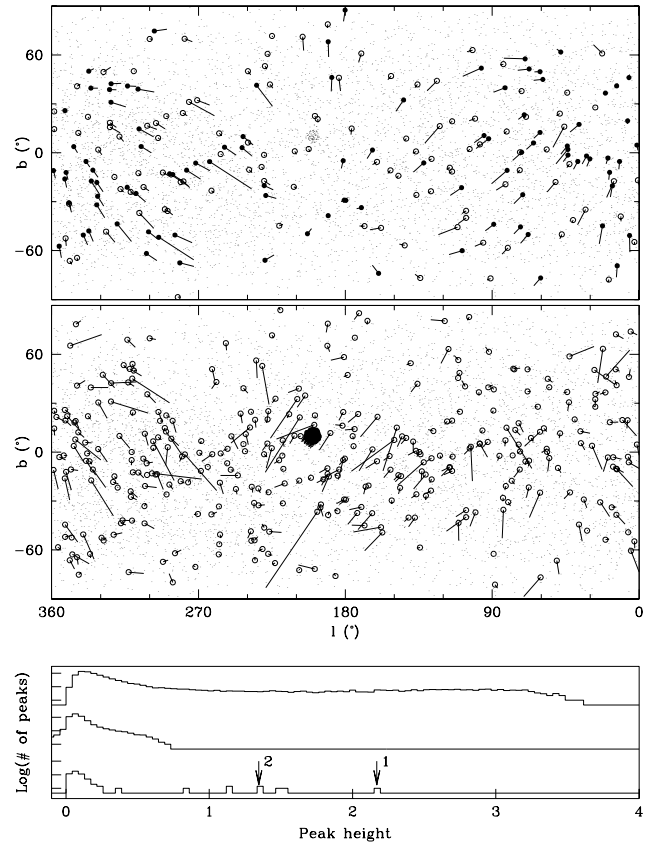,width=8.5truecm} 
  \caption{The bottom histogram in the bottom panel shows the
  significant peaks, indicated by the arrows, corresponding to the
  all-sky (1) and the confined (2) moving group present in this
  particular synthetic data set. The top and middle histogram are the
  Monte Carlo peak distribution before and after background
  subtraction (\S \ref{sec:peaks}). The ticks on the ordinate are
  spaced every two decades. The top panel shows the selected stars of
  the all-sky moving group, filled circles are genuine members and
  open circles are selected field stars. The small dots are the
  remaining stars in the data set. The middle panel is identical to
  the top panel but now shows the results of the second moving
  group. 95 genuine members are located around $(\ell,b) =
  (200^\circ\!,10^\circ\!)$.}
\label{fig:multiple_groups}
\end{figure}

We also apply the Spaghetti method to a synthetic data set which
contains two moving groups. The data set consists of 5000 field stars,
one group of 100 stars at 200~pc in the direction of $(\ell,b) =
(200^\circ\!,10^\circ\!)$ with a radius of 15~pc, and a group of 100
stars covering the whole sky, i.e., a distance of 20~pc in the
direction $(\ell,b) = (0^\circ\!,0^\circ\!)$ and a radius of
100~pc. Both moving groups have a small streaming velocity and a
velocity dispersion of 2~km~s$^{-1}$.  The field stars are generated
using the procedure described in \S \ref{sec:peaks}. We find several
significant peaks for this data set (Fig.\ \ref{fig:multiple_groups},
bottom panel), using $\sigma_{\rm int} = 2$~km~s$^{-1}$. The most
significant peak corresponds to the all-sky group while one of the
other peaks coincides with the velocity of the second moving group,
indicated with 1 and 2 in Fig.\ \ref{fig:multiple_groups},
respectively. Both peak velocities agree within 0.5~km~s$^{-1}$ with
those of the moving groups. The other significant peaks are phantom
peaks generated by both groups (cf. \S \ref{sec:results_peaks}). Using all
stars between 50 and 150~pc, we obtain $\sigma_{\rm median} =
4.20$~km~s$^{-1}$, resulting in the selection of 211 stars as member
of the all-sky group (Fig.\ \ref{fig:multiple_groups}, top panel). Of
these, 98 are classified correctly. We select 446 members for the
other group ($\sigma_{\rm median} = 7.02$~km~s$^{-1}$ for all stars
between 150 and 250~pc) of which 95 are genuine members (Fig.\
\ref{fig:multiple_groups}, middle panel).

In conclusion, the Spaghetti method succeeds in the detection of both
moving groups and in the selection of more than 90~per~cent of their
members. The only drawback, caused by the large field of view, are the
large numbers of misidentified members: 113/211 for the all-sky group
and 351/446 for the other. The large fraction of selected field stars
will make any analysis of all-sky moving groups extremely difficult;
only accurate age and chemical abundance information may shed further
light on the physical existence of such groups. However, the method
does reduce the number of stars of interest considerably, which makes
follow-up studies feasible.

\newdimen\digitwidth
\setbox0=\hbox{\rm0}
\digitwidth=\wd0
\catcode`?=\active
\def?{\kern\digitwidth}
\begin{center}
\begin{table*}
\caption{Results of the member selection for the Hyades open cluster
by the Spaghetti method compared to Perryman et al.\ (1998; P98). The
table shows the number of stars considered as member by P98 for
different confidence intervals, 68.3, 95.4, and 99.7 per cent
(corresponding to the 1, 2, and 3$\sigma$ confidence intervals), for
all members and for the members within 10 and 20~pc of the cluster
centre of mass. The P98 results are compared to the members found by
the Spaghetti method, SP, and the difference between P98 and SP, P98 -
SP, and SP - P98 respectively.}
\begin{tabular}{rrrrrrrrrr} \hline
 & \multicolumn{3}{c}{$\leq 3\sigma$} &
   \multicolumn{3}{c}{$\leq 2\sigma$} &
   \multicolumn{3}{c}{$\leq 1\sigma$}\\ 
\medskip
 & ???Total & r $<$ 10~pc & r $<$ 20~pc & 
   ???Total & r $<$ 10~pc & r $<$ 20~pc & 
   ???Total & r $<$ 10~pc & r $<$ 20~pc \\ 
P98      & 218 & 134 & 180 & 190 & 131 & 170 & 162 & 121 & 150 \\
P98 - SP &  56 &  11 &  29 &  30 &   8 &  19 &   7 &   0 &   3 \\ 
SP       & 168 & 124 & 154 & 168 & 124 & 154 & 168 & 124 & 154 \\
SP - P98 &   6 &   1 &   3 &   8 &   1 &   3 &  13 &   3 &   7 \\ \hline
\label{tab:hyades}
\end{tabular}
\end{table*}
\end{center}

\section{Tests on the Hyades and IC2602}\label{sec:hyades_ic2602}

Below, we discuss results obtained by the Spaghetti method for two
moving groups: the well-studied Hyades open cluster, and the open
cluster IC2602. Another application of the Spaghetti method, in
combination with the refurbished convergent point method of de Bruijne
(1999), can be found in de Zeeuw et al.\ (1999), who used the method
to improve and extend the membership lists for 12 nearby OB
associations.

\subsection{The Hyades}\label{sec:hyades}

Ever since its discovery as a moving group, the Hyades open cluster
has been at the centre of attention in studies of open cluster evolution,
distance calibration, and stellar evolution models. Here we
compare the most recent membership determination by Perryman et al.\
(1998, hereafter P98) to stars selected by the Spaghetti method.

\begin{figure*}
  \psfig{file=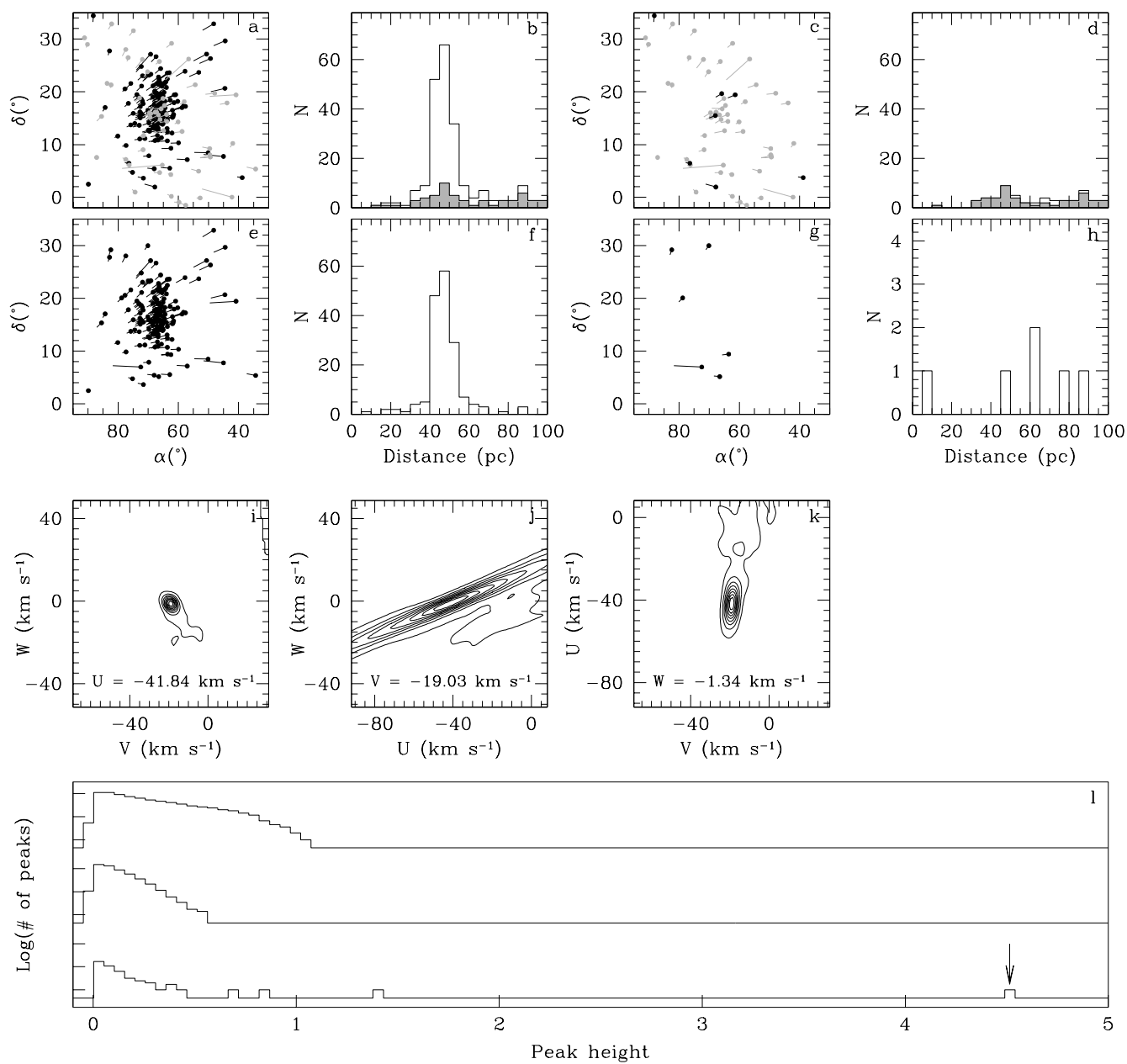,width=\textwidth} \caption{Panel
  (a) shows the positions and proper motions of the Hyades members as
  selected by Perryman et al.\ (1998; P98). The most significant
  members (within the 68.3 per cent confidence level) of P98 are
  denoted by black dots. The remaining stars lying within the 68.3 per
  cent and 99.73 per cent confidence level are denoted by grey
  dots. Panel (b) shows a histogram of the distances for all P98
  members. The grey histogram indicates the less significant
  members. The members with low significance are spatially less
  concentrated than the more significant members. Panel (c) shows the
  positions and proper motions of the P98 members which are not
  selected by the Spaghetti method; most of them have low significance
  and are located outside the main cluster, see the distance histogram
  in Panel (d). The colour scheme is the same as in Panel (a). Panel
  (e) shows the positions and proper motions of the Hyades members
  selected by the Spaghetti method. The corresponding distance
  histogram is shown in Panel (f). The positions and proper motions of
  the stars selected as member by the Spaghetti method which are not a
  P98 member are shown in Panel (g). Panel (h) shows the distance
  distribution of these stars. The Panels (i), (j), and (k) show
  slices through the SDF for the peak corresponding to the Hyades open
  cluster. The peak velocity is $(U,V,W) = (-41.84,-19.03,-1.34)~{\rm
  km}~{\rm s}^{-1}$, where $(U,V,W)$ are the Galactic Cartesian
  velocity coordinates. Panels (i) and (k) show a clear peak and a
  broader feature centred on the Solar motion. The elongated peak in
  Panel (j), stretched in the radial direction, is caused by the
  parallelism of the individual spaghettis. In general this effect results in a
  large uncertainty on the predicted radial velocity for the moving
  group. However, due to the large solid angle subtended by the Hyades
  on the sky and its significant peculiar velocity this is not a
  problem for the Hyades.  Panel (l) shows the significance of the
  peak created by the Hyades in the SDF: the top histogram shows the
  peak distribution for a Monte Carlo simulation as described in \S
  \ref{sec:peaks}, the second histogram is the peak height distribution for the
  same simulation but after the SDF background has been
  subtracted. The bottom histogram shows the peak height distribution
  for the Hyades data sample after subtraction of the Monte Carlo
  background. The ticks on the ordinate are spaced every two
  decades. The peak corresponding to the Hyades moving group is
  indicated with an arrow.}
\label{fig:hyades}
\end{figure*}

P98 base membership of the Hyades on the three-dimensional
velocity of a star and the three-dimensional centre-of-mass motion of
the cluster. A star is considered as member if these two are
consistent within their associated statistical errors within the 99.73
per cent confidence level. For stars with unknown radial velocity
membership is based on the tangential velocity only. P98 find 218
members of which 21 do not have a measured radial velocity. 134 of the
218 are located within 10~pc of the cluster centre of mass and 180
within 20~pc. The tidal radius of the cluster is 9~pc. P98 find a
centre-of-mass velocity, with respect to the Sun for the 134 
members within 10~pc, of $(U,V,W) =
(-41.70\pm0.16,-19.23\pm0.11,-1.08\pm0.11)$~${\rm km}~{\rm s}^{-1}$
and a distance of $46.34\pm0.27$~pc.

We apply the Spaghetti method to all Hipparcos stars in the field
$2^{\rm h}15^{\rm m} < \alpha < 6^{\rm h} 5^{\rm m}$ and $-2^\circ\! <
\delta < 35^\circ\!$ with parallaxes larger than 10~mas. This is the
same sample used by P98. We take the internal velocity dispersion,
$\sigma_{\rm int}$, to be 2.0~${\rm km}~{\rm s}^{-1}$. This value is
larger than the 0.3~${\rm km}~{\rm s}^{-1}$ found by P98. However,
when combined with the median error (1.64~${\rm km}~{\rm s}^{-1}$) for
stars between 25 to 65 pc from the Sun, this results in a radius for
the sphere of interest of $R_s = 2.58~{\rm km}~{\rm s}^{-1}$, 
similar to the $1\sigma$ errors in the centre-of-mass motion for the
Hyades (eq.~17 and fig.~16 of P98). 

The Hyades cluster generates a significant peak in the SDF (see
Fig.~\ref{fig:hyades} panels [i]--[l]). The peak stands well clear
from the background peak distribution expected for a similar data set
without moving groups (see \S\S \ref{sec:peaks} and
\ref{sec:results_peaks}). The other 3 significant peaks in
Fig.~\ref{fig:hyades} panel [l] are also generated by the
Hyades. These are random peaks superposed on that of the cluster. We
find a velocity for the Hyades from the position of the peak in the
SDF of $(U,V,W) = (-41.84,-19.03,-1.34)$~${\rm km}~{\rm s}^{-1}$ with
respect to the Sun. This is within 0.3~${\rm km}~{\rm s}^{-1}$ equal
to the velocity found by P98; their result for stars within 10~pc of
the cluster centre.

\begin{figure*}
  \psfig{file=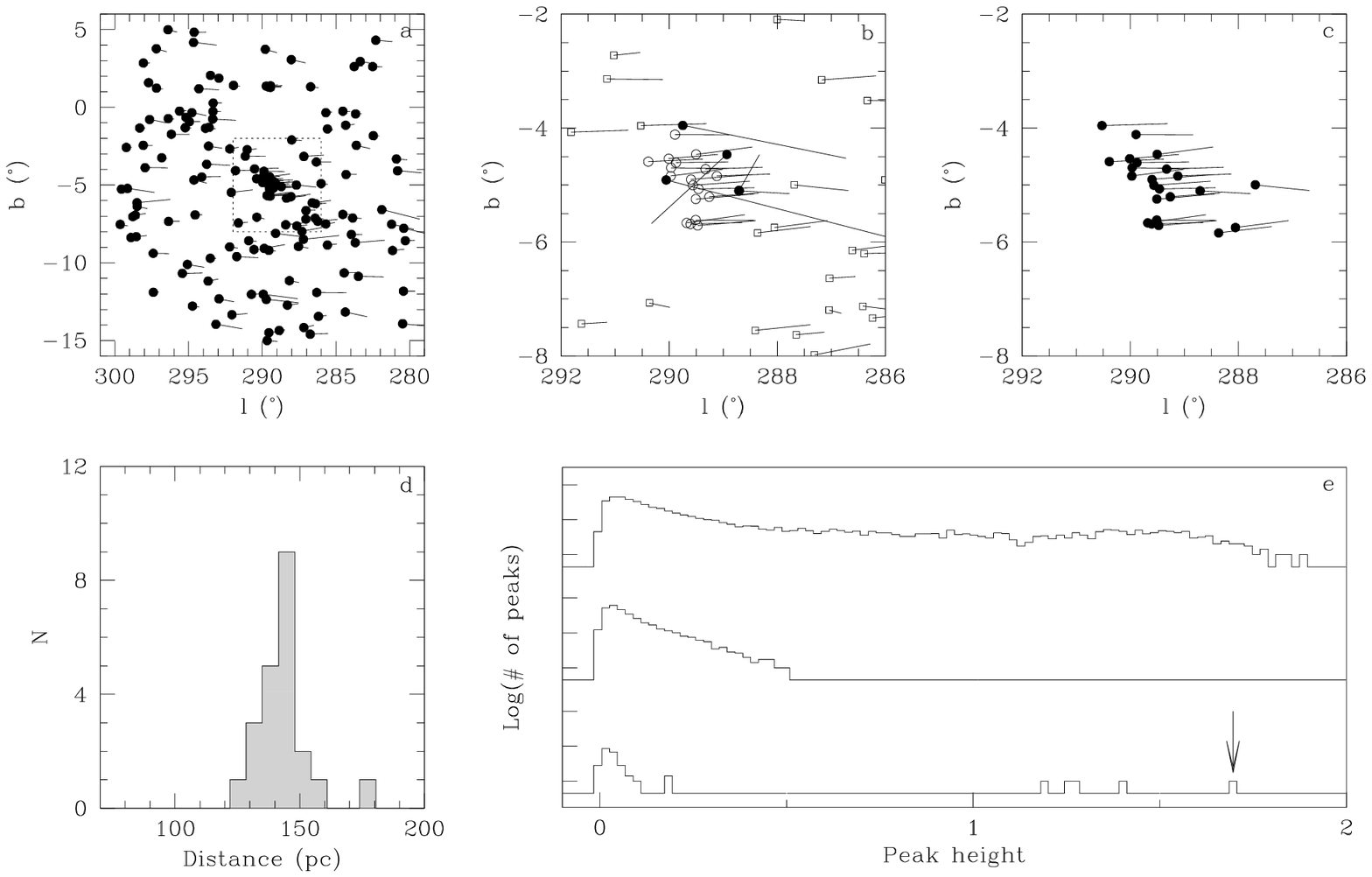,width=\textwidth} 
  \caption{Panel (a): Positions and proper motions of 156 stars
  classified as member of IC2602. Besides the clearly visible cluster
  at $(\ell,b) = (289^\circ\!,-5^\circ\!)$ the Spaghetti method also
  selects a large number of field stars. Panel (b) shows a
  $6^\circ\!\times6^\circ\!$ field centred on IC2602, also indicated
  by the dashed lines in panel (a). The open symbols show the 42 stars
  selected by the Spaghetti method. The 19 open circles are also in
  the membership lists of Whiteoak (1961) and Braes (1962). The 4
  filled circles are Whiteoak and Braes members, not confirmed by the
  Spaghetti method. The stars located on the edge of this field 
  (square dots) are, all but 4, considered as field stars. The
  resulting 23 IC2602 members are shown in panel (c). Panel (d)
  displays a distance histogram for these stars. Panel (e) shows the
  peak significance of IC2602 (arrow) as in Fig.~\ref{fig:hyades}.}
\label{fig:ic2602}
\end{figure*}

Using a membership threshold of $S_{\rm min} = 0.1$ (see \S
\ref{sec:results_membership}) we select 168 stars as belonging to the
Hyades. Of these, 6 are not in the member list of P98 and 56 of the
218 P98 members are not selected by the Spaghetti method (see also
Table~\ref{tab:hyades}). A comparison of the position, proper motion,
and parallax distribution for both approaches is presented in
Fig.~\ref{fig:hyades}. The 6 stars selected by the Spaghetti method
but not by P98 do not show any clustering in position on the sky and
in distance; and are most likely all field stars. Only one of these 6
stars (HIP20693) is listed as a classical member of the Hyades and was
rejected by P98 on the basis of its radial velocity. Most of the 56
P98 members not found by the Spaghetti method have a membership
significance outside the 68.3 per cent confidence level (table~2 in
P98) and are mostly located further than 20~pc from the centre of mass
(see Table~\ref{tab:hyades} and Fig.~\ref{fig:hyades}).

In conclusion, the Spaghetti method selects the majority of the P98
Hyades members, where the number of Spaghetti members missed by P98 is
less than the P98 members missed by the Spaghetti method. The
difference in membership lists mostly concerns stars having a low
significance in the P98 analysis.

\subsection{IC2602}\label{sec:ic2602}

IC2602 is a young open cluster centred on the second magnitude star
$\theta$~Car. Membership for the brightest stars, of spectral types A0
and earlier, was determined by Whiteoak (1961) and Braes (1962)
resulting in a total of 46 candidate members, at a distance of
$\sim$155~pc. Only recently, using photometric, spectroscopic, and
X-ray information, has it become possible to assign membership for
fainter stars, $V \sim 10$ to $\sim$18 mag. This resulted in about 50
additional members (see e.g., Prosser, Randich \& Stauffer 1996;
Stauffer et al.\ 1997; Foster et al.\ 1997; Randich et al.\ 1997). The
age estimates range from 8 Myr, based on the ages of the brightest
members (Whiteoak 1961; Braes 1962), to 30 Myr, based on isochrone
fits to the low mass members (Stauffer et al.\ 1997). 
As most of the fainter members of IC2602 are not listed in the
Hipparcos Catalogue we compare our results to those of Whiteoak and
Braes. Only 24 of their 46 members are contained in the Hipparcos
Catalogue. 

We apply the Spaghetti method to all stars in the Hipparcos Catalogue
in the field $280^\circ\!< \ell < 300^\circ\!$ and $ 5^\circ\!< b <
-15^\circ\!$ centred on IC2602. This field also contains part of the
Lower Centaurus Crux subgroup of the Sco OB2 association at
$118\pm2$~pc (de Zeeuw et al.\ 1999). These authors showed that
HIP52357, proposed as member of IC2602 by both Whiteoak (1961) and
Braes (1962), is a member of this association. We exclude this star in
the following discussion. Of the 1958 stars in the field 104 have $\pi
\leq 0$ and are not included in the selection. We use all stars with
distance between 50 and 250~pc to obtain a value of $\sigma_{\rm
median}$ (2.9~km~s$^{-1}$), and we adopt 2~${\rm km}~{\rm s}^{-1}$ for
the internal dispersion, $\sigma_{\rm int}$. The method finds 156
stars related to a significant peak in the SDF (see
Fig.~\ref{fig:ic2602} panels [a] and [e]) at a velocity $(U,V,W) =
(-12.6,-7.2,-0.5)~{\rm km}~{\rm s}^{-1}$. Due to the large field we
select a considerable number of field stars having indistinguishable
kinematics from the moving group (Fig.~\ref{fig:ic2602} panel
[a]). However, IC2602 is clearly visible. Note that the Spaghetti
method picks up a moving group of $\sim$30 stars out of 1854. In the
following we only consider the stars in a $6^\circ\!  \times
6^\circ\!$ field centred on $(\ell,b) = (289^\circ\!,-5^\circ\!)$.

\begin{table}
\caption{Hipparcos identifiers and $S$ values for the 23 IC2602 members}
\begin{center}
\begin{tabular}{rrrrrrrr} \hline
\medskip
HIP&$S$&HIP&$S$&HIP&$S$&HIP&$S$\\
51131&0.67& 51203& 0.66& 51300& 0.42& 51794& 0.23\\
52059&0.78& 52116& 0.60& 52132& 0.86& 52160& 0.89\\
52171&0.48& 52221& 0.68& 52261& 0.80& 52293& 0.90\\
52328&0.88& 52370& 0.88& 52419& 0.82& 52502& 0.71\\
52678&0.67& 52701& 0.68& 52736& 0.93& 52815& 0.64\\
52867&0.40& 53016& 0.84& 53330& 0.42&      &     \\ \hline    
\end{tabular}
\end{center}
\label{tab:ic2602}
\end{table}

We select 42 stars in this field of which 19 are also in the lists of
Whiteoak and Braes (Fig.~\ref{fig:ic2602} panel [b]). Only 4 of the
Whiteoak and Braes members are not selected by the Spaghetti
method. The astrometric parameters for one of these, HIP517979, are a
solution for the B and C component of the triple system
CCDM~10350-6408 of which component A, HIP51794, is selected as member
of IC2602. The complex configuration of this triple system might have
caused the Hipparcos solution to be in error as is also indicated by
fields H29 and H30 in the Hipparcos Catalogue. The other three stars,
HIP52178, HIP52216, and HIP52839, all have good solutions. The
remaining 23 stars are almost all situated near the edge of the field
and at distances larger than 200~pc. Only four of these stars,
HIP51131, HIP51202, HIP51300, and HIP53330, the ones closest to the
cluster, are considered as genuine members of IC2602. We consider
these four, plus all stars selected by Whiteoak (1961), Braes(1962) as
well as the Spaghetti method --- 23 stars in total --- as genuine IC2602
members (see Table~\ref{tab:ic2602} and Fig.~\ref{fig:ic2602} panel [c]).

Although we did not use the parallaxes directly in the selection
process the distance distribution of the IC2602 members shows a sharp
peak, indicating that the cluster is not only confined in velocity
space but also in configuration space (Fig.~\ref{fig:ic2602} panel
[d]). We find the mean distance of the 23 cluster members, using the
mean parallax of the cluster and corrected for the known biases as
described in de Zeeuw et al.\ (1999), to be $145\pm4$~pc. This is
$\sim$10~pc closer than the previous estimates of Whiteoak and Braes.

\section{Summary and discussion}\label{sec:summary}

We have presented a new method to identify moving groups, and select
their members, based on position, proper motion, and parallax information
only, i.e., using the Hipparcos Catalogue. The astrometric parameters
and their errors define a probability distribution function
represented by a cylinder with an elliptical cross section in
velocity space. Moving group stars, having the same spatial velocity, 
produce an overdensity in the combined probability function, the SDF,
in velocity space. Assessing the significance of these peaks allows
for the detection of moving groups and hence the selection of members.

The characteristics of the method have been tested on synthetic
data. Typically the Spaghetti method finds $\sim$85 per cent of the
synthetic cluster members for distances out to 600~pc. The
contamination by field stars increases rapidly after 450~pc which
makes cluster detection impossible beyond $\sim$750~pc. This is due to
the typical error on the Hipparcos parallaxes, $\sim$1~mas,
corresponding to errors in the tangential velocity of
$\sim$20~km~s$^{-1}$ at 750~pc, which is of the same order as the
structure in velocity space. The method has primarily been
developed to identify moving groups in the Hipparcos database. Any
further analysis should take into account the selection effects and
kinematic biases which originate from the construction of the
Catalogue. For example, the Catalogue is biased towards high proper
motion stars although it has a formal absolute proper motion cut off
of zero.

Results were presented for the Hyades and IC2602 open clusters. For
the Hyades we find most of the members proposed by Perryman et al.\
(1998). The differences between the two membership lists generally
concern P98 members of low significance, most likely field
stars. Furthermore, we confirm most of the bright IC2602 members as
found by Whiteoak (1961) and Braes (1962), and add four new ones.

The typical errors of $\sim$1~mas in proper motion and parallax
prevent identification of moving groups in the Hipparcos Catalogue
more distant than $\sim$750~pc from the Sun. The future space
astrometry mission GAIA (see e.g., Lindegren \& Perryman 1996) aims at
observing positions, parallaxes, and proper motions with accuracies of
$10~\mu$as and $10~\mu{\rm as}~{\rm yr}^{-1}$, respectively, at $V
\sim 15$~mag, allowing the detection of moving groups out to 70~kpc.

The Spaghetti method can in principle be used to search for
kinematic substructure in the Galactic Halo caused by infalling
satellites (e.g., Lynden--Bell 1976; Lynden--Bell \& Lynden--Bell
1995). However, the Hipparcos Catalogue is far from complete for these
stars which makes the interpretation of the existence of real moving
groups in the halo extremely difficult (e.g., Aguilar \&
Hoogerwerf 1998).

\bigskip
It is a pleasure to thank Anthony Brown, Jos de Bruijne, Michael
Perryman, Tim de Zeeuw, and HongSheng Zhao for helpful discussions
and/or careful reading of the manuscript. We also thank the
anonymous referee. This research was funded in part by DGAPA grant
N114496. We thank the IAUNAM/Ensenada for access to their Origin-2000
computer. L.\ Aguilar wishes to thank Sterrewacht Leiden for
hospitality during a sabbatical visit, and NWO, the Netherlands
Organization for Scientific Research, for a Bezoekersbeurs which
supported this visit.

\bsp

\end{document}